\documentclass[aip,reprint,graphicx,twocolumn,superscriptaddress,nofootinbib]{revtex4-2}
\usepackage{amssymb,amsmath,amsbsy,graphicx,xcolor,times,subfigure,marvosym,color,bm,multirow,epstopdf,color,soul}
\begin{document}




\title{Quantum Enhanced Probes of Magnetic Circular Dichroism}

\author{Chengyun Hua}
\email[]{huac@ornl.gov}
\affiliation{Materials Science and Technology Division, Oak Ridge National Laboratory, Oak Ridge TN 37831 USA}
\author{Claire E. Marvinney}
\affiliation{Computational Sciences and Engineering Division, Oak Ridge National Laboratory, Oak Ridge TN 37831 USA}
\author{Seongjin Hong}
\affiliation{Physics Division, Oak Ridge National Laboratory, Oak Ridge TN 37831, USA}
\author{Matthew Feldman}
\affiliation{Computational Sciences and Engineering Division, Oak Ridge National Laboratory, Oak Ridge TN 37831 USA}
\author{Yun-Yi Pai}
\affiliation{Materials Science and Technology Division, Oak Ridge National Laboratory, Oak Ridge TN 37831 USA}
\author{Michael Chilcote}
\affiliation{Materials Science and Technology Division, Oak Ridge National Laboratory, Oak Ridge TN 37831 USA}
\author{Joshua Rabinowitz}
\affiliation{Department of Physics, Duke University, Durham, North Carolina 27708, USA}
\author{Raphael C. Pooser}
\affiliation{Computational Sciences and Engineering Division, Oak Ridge National Laboratory, Oak Ridge TN 37831 USA}
\author{Alberto Marino}
\affiliation{Computational Sciences and Engineering Division, Oak Ridge National Laboratory, Oak Ridge TN 37831 USA}
\author{Benjamin J. Lawrie}
\email[]{lawriebj@ornl.gov; This manuscript has been authored by UT-Battelle, LLC, under contract DE-AC05-00OR22725 with the US Department of Energy (DOE). The US government retains and the publisher, by accepting the article for publication, acknowledges that the US government retains a nonexclusive, paid-up, irrevocable, worldwide license to publish or reproduce the published form of this manuscript, or allow others to do so, for US government purposes. DOE will provide public access to these results of federally sponsored research in accordance with the DOE Public Access Plan (http://energy.gov/downloads/doe-public-access-plan).}
\affiliation{Materials Science and Technology Division, Oak Ridge National Laboratory, Oak Ridge TN 37831 USA}

\date{\today}

\begin{abstract}
 Magneto-optical microscopies, including optical measurements of magnetic circular dichroism, are increasingly ubiquitous tools for probing spin-orbit coupling, charge-carrier g-factors, and chiral excitations in matter, but the minimum detectable signal in classical magnetic circular dichroism measurements is fundamentally limited by the shot-noise limit of the optical readout field. Here, we use a two-mode squeezed light source to improve the minimum detectable signal in magnetic circular dichroism measurements by 3 dB compared with state-of-the-art classical measurements, even with relatively lossy samples like terbium gallium garnet. We also identify additional opportunities for improvement in quantum-enhanced magneto-optical microscopies, and we demonstrate the importance of these approaches for environmentally sensitive materials and for low temperature measurements where increased optical power can introduce unacceptable thermal perturbations.
\end{abstract}


\maketitle 

\section{Introduction}
Magnetic circular dichroism (MCD) has become a crucial tool for probing electronic band structure, electron g-factors, spin-orbit coupling, crystal-field interactions and magnetic and electronic phase transitions in quantum heterostructures~\cite{song2018giant,xia2022magnetic,scott1975magnetic,foglia2022going,meeker2015high,chen2021strong,pham2022strong,lin2022field}. Many magneto-optical spectroscopies are still performed with excessive classical noise, and state-of-the-art classical measurements are still limited by the photon shot-noise limit (SNL), where the minimum resolvable signal scales with the square root of the number of photons in the measurement~\cite{xia2006modified,gomez2020high,crooker2004spectroscopy}. The SNL defines the fundamental limit of detection for a classical light source for a given laser power and measurement time. In some cases, MCD signals can be strongly enhanced when the optical readout field is resonant with electronic transitions in matter~\cite{chen2021strong}, and the sensitivity of any optically transduced measurement can always be improved with increased laser power or increased measurement times. Unfortunately, increased laser power can also introduce substantial unwanted thermal perturbations to the measurement, and increased measurement times are a substantial obstacle to measurements of out-of-equilibrium processes. When resonant enhancement, increased laser power, and increased measurement times aren't possible, or when they don't provide sufficient measurement sensitivity, the obvious next step is to further suppress measurement noise. Balanced photodetection schemes can be used to suppress classical noise sources up to the common-mode rejection ratio of the detector, but the shot noise on a coherent optical field will add in quadrature in a balanced photodetection measurement.

Squeezed states of light suppress the uncertainty in a given quadrature compared with the SNL (at the expense of increased uncertainty in another quadrature). Decades after they were first proposed as resources for quantum enhanced detection of gravitational waves\cite{Caves}, squeezing has increasingly emerged as a practical resource for quantum sensing and quantum metrology~\cite{taylor2013biological,taylor2014subdiffraction,harms2003squeezed,kimble2001conversion,pooser_ultrasensitive_2015,dowran2018quantum,aasi2013enhanced,lawrie_extraordinary_2013,fan2015quantum,hoff_quantum-enhanced_2013,pooser_ultrasensitive_2015,pooser2020truncated,otterstrom_nonlinear_2014,lawrie2020squeezing,lawrie2019quantum,xu2022stimulated,michael2019squeezing,de2020quantum,li2022quantum,prajapati2021quantum}. Reducing the noise in a given quadrature with squeezed states of light provides an improved minimum detectable signal for constant optical power and measurement time. Squeezed quantum sensors relying on absorption and phase shifts in plasmonic media increasingly offer improved sensitivity compared with classical plasmonic sensors~\cite{pooser_plasmonic_2015, lawrie_extraordinary_2013, fan2015quantum,lee2021quantum, dowran2018quantum}, and squeezed states of light have been used to provide quantum enhanced sensitivity in bioimaging~\cite{taylor2013biological,taylor2014subdiffraction}, Raman spectroscopy~\cite{xu2022stimulated,michael2019squeezing,de2020quantum}, Brillouin spectroscopy\cite{li2022quantum}, and two-photon spectroscopies~\cite{prajapati2021quantum}. Experimental reports of quantum enhanced beam displacement measurements with squeezed light date back 20 years~\cite{treps2003quantum}, and squeezed beam displacement measurements have since been used for applications spanning atomic force microscopy and optomechanical magnetometry ~\cite{pooser_ultrasensitive_2015,pooser2020truncated,schafermeier2016quantum} to gravitational wave detection ~\cite{harms2003squeezed, kimble2001conversion, aasi2013enhanced}.  However, despite a recent proposal illustrating a path to quantum-enhanced magneto-optical Kerr effect measurements~\cite{pai2022magneto}, squeezed light has still not been used to suppress the noise in magneto-optical spectroscopies in condensed matter physics, despite the substantial opportunities associated with probing environmentally sensitive quantum states in emerging quantum materials with reduced laser powers and quantum enhanced measurement sensitivity~\cite{song2018giant,xia2022magnetic,scott1975magnetic,foglia2022going,meeker2015high,chen2021strong,pham2022strong,lin2022field}.

\section{Approach}

\begin{figure}[htbp]
\includegraphics[scale = 0.45]{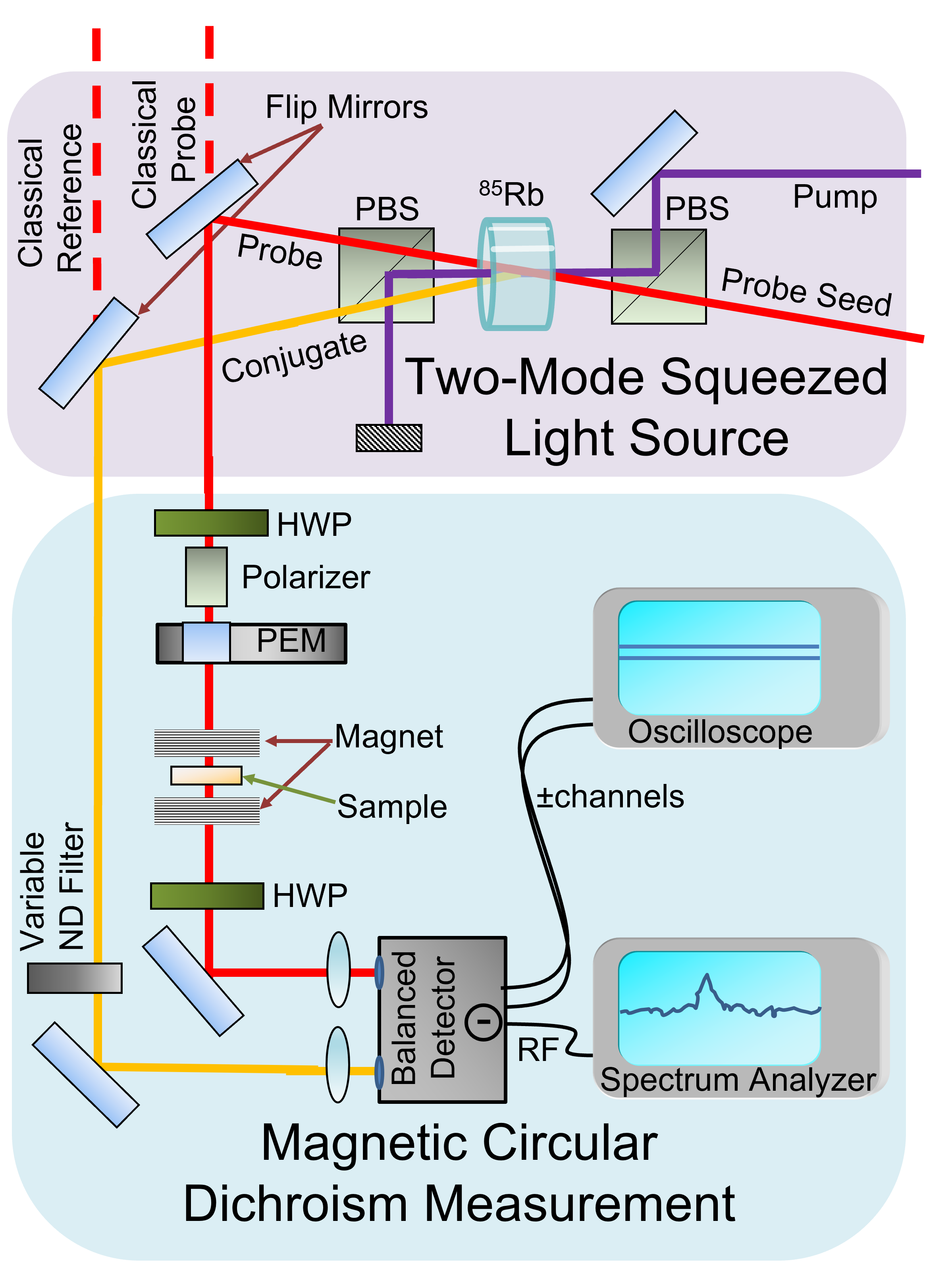}
\caption{Schematic of quantum enhanced MCD experiment.  The two-mode squeezed light source is generated via four-wave mixing in \textsuperscript{85}Rb vapor before the MCD signal is transduced onto the probe field and a balanced photodetection scheme is used to suppress noise below the SNL. PBS, polarizing beam splitter; PEM, photoelastic modulator; BD, balanced detector; HWP, half-wave plate. Two flip mirrors are used to convert between classical and squeezed measurements. The sample is mounted in a variable temperature cryostat in a uniaxial electromagnet, though all data reported here was acquired at room temperature.} 
\label{Figure1}
\end{figure}

Here, we demonstrate 3 dB improvement in the minimum resolvable signal in circular dichroism measurements of lossy samples using a two-mode squeezed state of light generated via four-wave mixing (FWM) in \textsuperscript{85}Rb vapor as illustrated schematically in Figure~\ref{Figure1}. The pump and probe seed beams are both derived from the same continuous-wave Ti:sapphire laser, and as shown in Figure~\ref{Figure1}, an acousto-optic modulator is used to redshift the probe seed beam by 3.039 GHz from the pump before the two beams are weakly focused and mixed at an angle of 0.32$^{\circ}$ in a 12.7 mm long $^{85}$Rb vapor cell held at 100.3$^{\circ}$C. The four-wave mixing process generates probe and conjugate fields that exhibit intensity difference and quadrature squeezing, though the measurements described here benefit from the simplicity of intensity difference measurements. After the cell, a half-wave plate and a Glan-Taylor linear polarizer are used to polarize the probe field at 45$^{\circ}$ with respect to the horizontal axis with high polarization contrast and minimal optical loss. A photoelastic modulator (PEM; Hinds Instrument PEM-200) converts the linear polarization into left- and right-handed circular polarization alternately at a principal frequency ($\omega$) of around 50 kHz. Its retardance is set to one-fourth of the wavelength ($\lambda/4$) of the light such that the PEM acts as an oscillating quarter-wave plate. Following the PEM, a sample in a magnetic field oriented parallel to the direction of propagation of the probe produces the Faraday rotation $\theta_F = d\pi(n_r-n_l)/\lambda$ and Faraday ellipticity $\eta_F = d\pi(k_r-k_l)/\lambda$, where $\lambda$ is the wavelength of the light, $d$ is the thickness of the sample, $n_r$ and $n_l$ are the refractive indices for right- and left-handed circularly-polarized light, and $k_r$ and $k_l$ are the extinction coefficients for right- and left-handed circularly-polarized light. The conjugate beam serves as a reference for the measurement and is detected without interacting with either the PEM or the sample.  A variable neutral density filter is used to introduce balanced loss on the conjugate field and thus optimize the quantum noise reduction in the measurement. A second rotating half wave plate is used to minimize unwanted polarization rotation due to a combined effect of linear birefringence from all the optics and polarization sensitivity of the detector. A detailed theoretical analysis of this effect on the detected signal can be found elsewhere~\cite{cao2008reduction}. A balanced photodiode (Thorlabs PDB450A) with a gain setting of 10$^5$ is used to perform intensity difference measurements with both the two-mode squeezed state generated by four-wave mixing and the classical reference laser readout fields denoted classical probe and classical ref in Figure 1. The intensity difference signal is measured directly on a spectrum analyzer (Siglent SSA3021X) with a video bandwidth of 300 Hz and resolution bandwidth of 3000 Hz. 

\begin{figure}[hb!]
\centering
\includegraphics[scale = 0.45]{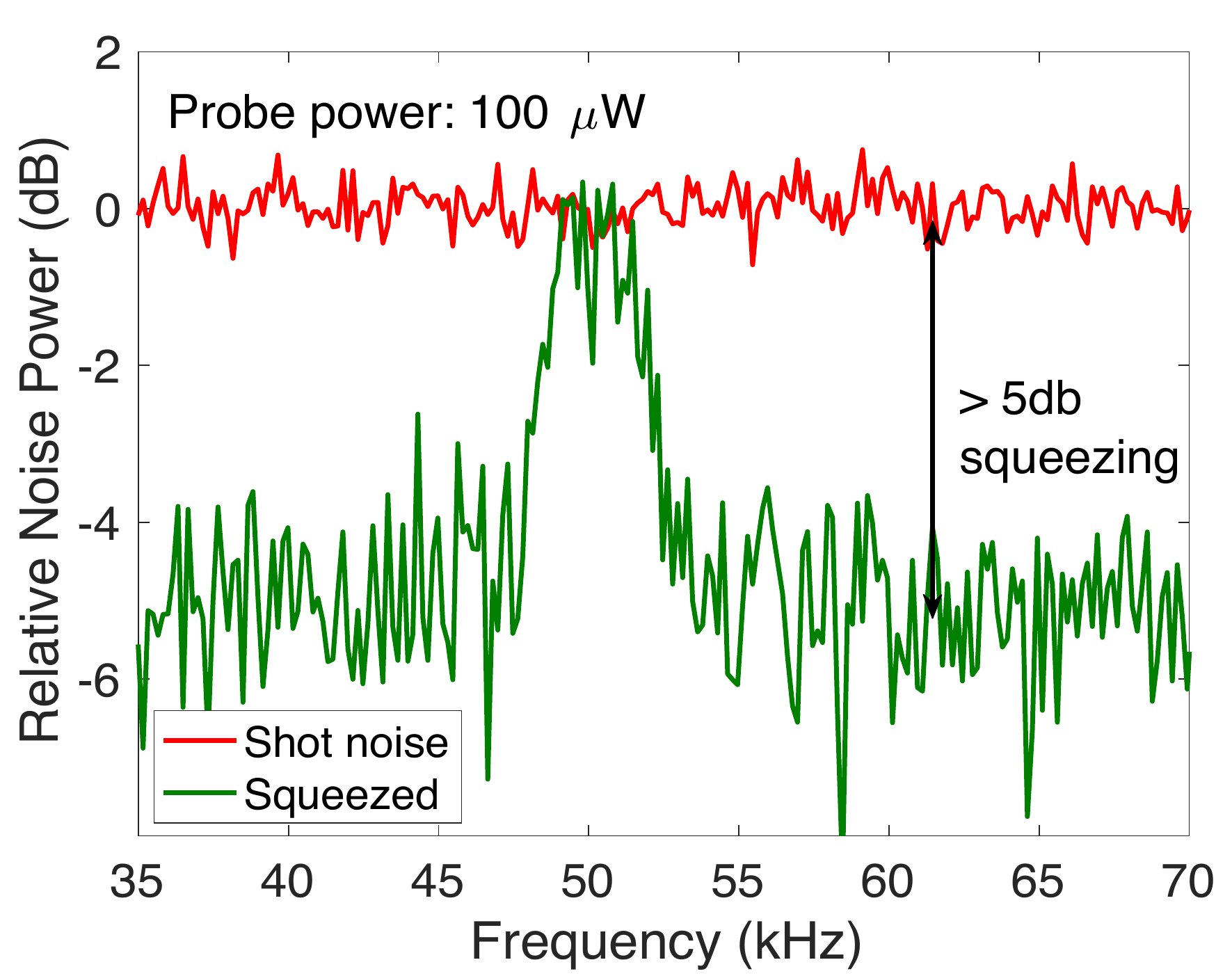}
\caption{Demonstration of a quantum-enhanced polarization sensitive signal (green) without a sample normalized against a shot noise limited trace (red) acquired with a probe power of 100 $\mu$W. In the absence of a sample, the signal detected at 50 KHz is due to a combined effect of linear birefringence and photodiode polarization sensitivity.} 
\label{Figure2}
\end{figure}

\begin{figure*}[htbp]
\centering
\includegraphics[scale = 0.25]{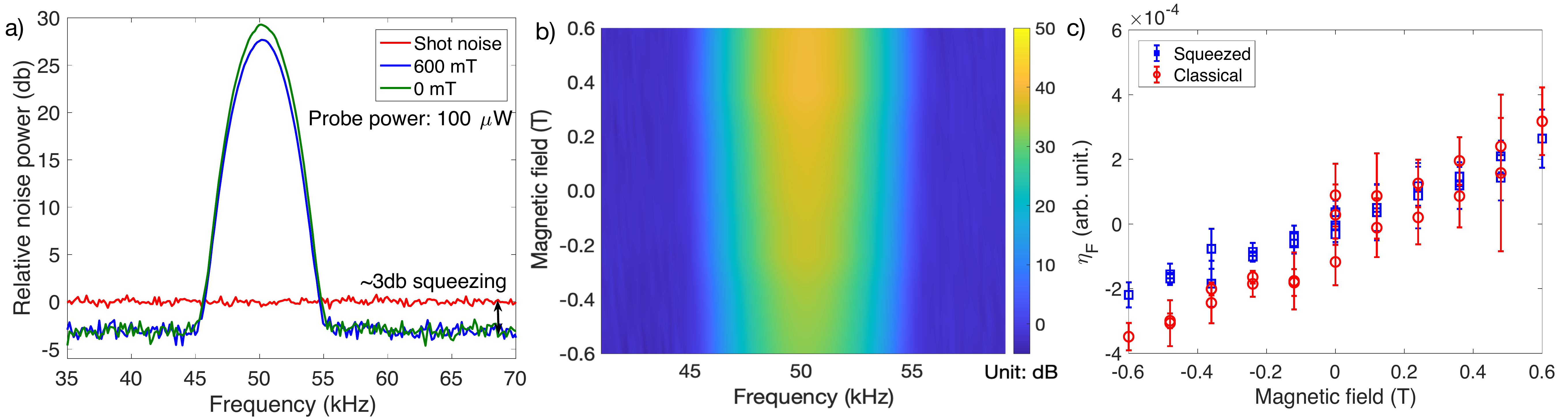}
\caption{Quantum enhanced MCD measurement. (a) SA traces around 50 kHz at 0 (green line) and 600 (blue line) mT compared to shot noise floor (red line) acquired with 100 $\mu$W of probe power. (b) Field-dependent SA traces. (c) Changes in ellipticity, $\eta_F$, of TGG as a function of magnetic field.} 
\label{Figure3}
\end{figure*}

In a simple classical MCD measurement, the probe light after the sample is directly sent to a photodiode and the optical power $P$ at the detector can easily be calculated using Jones matrices of optical elements in which the sample is considered as a circular dichroic retarder given by
\begin{equation}
P = P_0\left\{1-\text{sin}\left[\frac{\pi}{2}\text{sin}(\omega t)\right]\text{tanh}(2\eta_F)\right\},
\label{eq:CDSignal_singledetector}
\end{equation}
for incident optical power $P_0$ on the sample with the absorption due to each optical element neglected. The signal at the sideband with frequency $\omega$ is derived by Fourier analysis of Eq.~(\ref{eq:CDSignal_singledetector}) as 
\begin{equation}
P_{\omega}\text{sin}(\omega t) = \{-2P_0\text{tanh}[2\eta_F J_1(\pi/2)]\}\text{sin}(\omega t),\label{eq:Pomega}    
\end{equation}
where $J_1(\pi/2)$ is the first-order Bessel function evaluated at $\pi/2$. $P_\omega$ is typically measured with a lock-in amplifier or a spectrum analyzer (SA). Note that $P_\omega$ is independent of $\theta_F$, which is only contained in the second harmonic signal, $P_{2\omega}$. 

\section{Results}

Classical balanced photodetection schemes like the approach illustrated in Figure 1 using equal laser power for the classical probe and reference channels are often used to suppress excess classical noise sources and enable photon shot-noise-limited sensitivity. Such an approach retains the same signal amplitude as in Eq.~2. The quantum enhanced MCD measurement illustrated in Fig.~1 relies on quantum correlations between the probe and conjugate fields to suppress the noise floor below the SNL. As previously shown~\cite{pooser_plasmonic_2015}, the squeezed noise floor can then be defined - normalized to the SNL - as
\begin{equation}
\Delta N_{-}^2 = 1 - \frac{2\eta (G-1)}{2G-1} , \label{eq:3}
\end{equation}
where $N_-$ is the photon number difference, $\eta$ is the total transmission of the probe and the conjugate fields from the vapor cell to the detector, and $G$ is the gain of the four-wave mixing process in the vapor cell.  For all $\eta>0$ and $G>1$, the fundamental limit of detection is then suppressed below the SNL, but the quantum enhancement is suppressed with increasing optical loss.

As classical and quantum noise sources are suppressed, understanding background signals that can contaminate MCD measurements becomes increasingly critical.  To that end, we start here by characterizing the detected signal at 50 kHz in the absence of a sample. For these reference measurements, the total loss in the optics train after the vapor cell is dominated by the 5\% loss at the photodiode (due to 95\% detector efficiency) and we observe at least 5 dB quantum noise reduction relative to the SNL as shown in Fig.~\ref{Figure2} for measurements performed with 100~$\mu$W of probe power. Even with no sample, a substantial signal is observed at 50 kHz, which can be explained as a result of small birefringence in the optics shown in Fig.~\ref{Figure1} and a small polarization sensitivity in the balanced photodiode. The signal shown in Fig.~\ref{Figure2} was minimized (but not completely suppressed) by tuning the angle of the second half wave plate. Further suppression of this background signal below the SNL hinges on improved polarization stability, suppression of unwanted birefringence effects, and reduced photodiode polarization dependence.

With the background signal minimized, we then move on to measure the magnetic circular dichroism of a prototypical terbium gallium garnet (Tb$_3$Ga$_5$O$_{12}$, TGG) crystal with a thickness of 500~$\mu$m. As shown in Fig.~\ref{Figure3}(a), the 20\% attenuation introduced by the TGG crystal suppresses the measured squeezing to roughly 3~dB below the SNL, but substantial suppression of quantum noise still exists even in this limit of a relatively lossy sample. Fig.~\ref{Figure3}(a) illustrates SA traces normalized to the SNL acquired with magnetic fields of 0 and 600 mT and constant probe power at the detector of 100~$\mu$W. The substantial zero-field signal at 50 kHz is due to the circular dichroic retardance of TGG and to the other smaller birefringence and polarization dependent effects seen in the absence of a sample. The relative change in the 50 kHz signal as we vary the strength of the magnetic field contains the desired MCD information, as shown in Fig.~\ref{Figure3}(b). Using Eq.~(\ref{eq:Pomega}) and offsetting any ellipticity induced by TGG at 0~mT, we can convert the measured signal at 50~kHz to $\eta_F$ as a function of magnetic field. The calculated signals acquired with a squeezed readout field and a classical reference using a lock-in amplifier in lieu of the SA are plotted in Fig.~\ref{Figure3}(c), where we see that the error bars on the squeezed measurements are on average $35\%$ smaller than the error bars on the classical measurement. However, it is critical to understand that the noise on the measurement is determined by a combination of the noise floor in the optical readout field around 50 kHz (where quantum noise reduction is observed), fluctuations at 50 kHz that emerge due to low frequency noise that is transferred to the 50 kHz sideband by the PEM, and polarization instability in the optics train that can introduce additional unwanted signal at 50 kHz. Optimization of the measurement bandwidth can help to suppress the contribution of low frequency noise to the measurement sensitivity\cite{atkinson2021quantum}, and further optimization of the polarization control in the optics train can help to minimize the contribution of remaining classical noise sources at the 50 kHz carrier frequency.

\section{Conclusion}

Using a two-mode squeezed light source to suppress photon shot noise and reduce the minimum detectable signal in MCD measurements will enable improved measurement sensitivity, and critically, reduced laser power (and reduced photothermal perturbations) for cryogenic optical probes of emerging quantum materials\cite{pai2022magneto}. Samples with higher transmissivity will exhibit improved noise floors, but substantial suppression of the photon shot noise was still observed here even with a relatively lossy TGG sample. The excess zero-field signal seen in these proof-of-principle experiments limits the immediate benefit of this approach for general applications, but improved control of the polarization state throughout the optics train and optimization of the measurement bandwidth should suppress the unwanted zero-field signal and enable a true quantum advantage for squeezed MCD measurements.

\begin{acknowledgments}

This research was supported by the U. S. Department of Energy, Office of Science, Basic Energy Sciences, Materials Sciences and Engineering Division. The squeezed light source was developed and optimized with support from the U.S. Department of Energy, Office of Science, National Quantum Information Science Research Centers, Quantum Science Center.
\end{acknowledgments}

\section*{Data Availability Statement}
Data underlying the results presented in this paper may be obtained from the authors upon reasonable request.

%

\end{document}